# Beyond the exome: what's next in diagnostic testing for Mendelian conditions

Monica H. Wojcik[1,2,3], Chloe M. Reuter[4], Shruti Marwaha[4], Medhat Mahmoud[5], Michael H. Duyzend[1,2,6], Hayk Barseghyan[7,8], Bo Yuan[9], Philip M. Boone[1,2,6], Emily E. Groopman[1,2,6], Emmanuèle C. Délot[8,10,11], Deepti Jain[12], Alba Sanchis-Juan[1,6], Genomics Research to Elucidate the Genetics of Rare Diseases (GREGoR) Consortium, Lea M. Starita[13,14], Michael Talkowski[1,6,15,16], Stephen B. Montgomery[17,18,19], Michael J. Bamshad[13,14,20], Jessica X. Chong[13,20], Matthew T. Wheeler[4], Seth I. Berger[21], Anne O'Donnell-Luria[1,2,22], Fritz J. Sedlazeck[5,23], Danny E. Miller[13,20,24,*]

1. Program in Medical and Population Genetics, Broad Institute of MIT and Harvard, Cambridge, MA 02142 USA
2. Division of Genetics and Genomics, Boston Children's Hospital, Harvard Medical School, Boston, MA 02115 USA
3. Division of Newborn Medicine, Boston Children's Hospital, Harvard Medical School, Boston, MA, USA
4. Department of Medicine, Division of Cardiovascular Medicine, Stanford University School of Medicine, Stanford, CA 94305 USA
5. Human Genome Sequencing Center, Baylor College of Medicine, One Baylor Plaza, Houston TX 77030 USA
6. Center for Genomic Medicine, Massachusetts General Hospital, Boston, MA 02114 USA
7. Center for Genetics Medicine Research, Children's National Research Institute, Children's National Hospital, Washington, DC 20010 USA
8. Department of Genomics and Precision Medicine, School of Medicine and Health Sciences, George Washington University, Washington, DC 20037 USA
9. Department of Molecular and Human Genetics and Human Genome Sequencing Center, Baylor College of Medicine, One Baylor Plaza, Houston TX 77030 USA
10. Center for Genetics Medicine Research, Children's National Research and Innovation Campus, Washington, DC, USA
11. Department of Pediatrics, George Washington University, School of Medicine and Health Sciences, George Washington University, Washington, DC 20037 USA
12. Department of Biostatistics, School of Public Health, University of Washington, Seattle WA 98195 USA
13. Brotman Baty Institute for Precision Medicine, University of Washington, Seattle, WA 98195 USA
14. Department of Genome Sciences, University of Washington, Seattle, WA, 98195 USA
15. Department of Neurology, Massachusetts General Hospital and Harvard Medical School
16. Stanley Center for Psychiatric Research, Broad Institute of MIT and Harvard, Cambridge, MA
17. Department of Biomedical Data Science, Stanford University School of Medicine, Stanford, CA 94305 USA
18. Department of Genetics, Stanford University School of Medicine, Stanford, CA 94305 USA
19. Department of Pathology, Stanford University School of Medicine, Stanford, CA 94305 USA
20. Department of Pediatrics, Division of Genetic Medicine, University of Washington, Seattle, WA 98195 USA
21. Center for Genetics Medicine Research and Rare Disease Institute, Children's National Hospital, Washington, DC 20010 USA
22. Center for Genomic Medicine, Analytic and Translational Genetics Unit, Massachusetts General Hospital, Boston, MA 02114 USA
23. Department of Computer Science, Rice University, 6100 Main Street, Houston, TX, 77005 USA
24. Department of Laboratory Medicine and Pathology, University of Washington, Seattle, WA, 98195 USA

*To whom correspondence should be addressed: dm1@uw.edu

**ABSTRACT**

Despite advances in clinical genetic testing, including the introduction of exome sequencing (ES), more than 50% of individuals with a suspected Mendelian condition lack a precise molecular diagnosis. Clinical evaluation is increasingly undertaken by specialists outside of clinical genetics, often occurring in a tiered fashion and typically ending after ES. The current diagnostic rate reflects multiple factors, including technical limitations, incomplete understanding of variant pathogenicity, missing genotype–phenotype associations, complex gene–environment interactions, and reporting differences between clinical labs. Maintaining a clear understanding of the rapidly evolving landscape of diagnostic tests beyond ES, and their limitations, presents a challenge for non-genetics professionals. Newer tests, such as short-read genome or RNA sequencing, can be challenging to order and emerging technologies, such as optical genome mapping and long-read DNA or RNA sequencing, are not available clinically. Furthermore, there is no clear guidance on the next best steps after inconclusive evaluation. Here, we review why a clinical genetic evaluation may be negative, discuss questions to be asked in this setting, and provide a framework for further investigation, including the advantages and disadvantages of new approaches that are nascent in the clinical sphere. We present a guide for the next best steps after inconclusive molecular testing based upon phenotype and prior evaluation, including when to consider referral to a consortium such as GREGoR, which is focused on elucidating the underlying cause of rare unsolved genetic disorders.

**Introduction**

The evaluation of an individual with a suspected Mendelian condition begins with a careful physical examination, review of family history, and evaluation of existing laboratory data. Together, this information helps the ordering provider decide whether an individual would benefit from genetic testing and whether to order focused or broad testing. Broad approaches to genetic testing, such as exome sequencing (ES), have benefited from access to large collections of control data (gnomAD[1]) and catalogs of pathogenic variation (e.g., ClinVar[2]), investment in infrastructure to support clinical genetic services, and changes in payor policies in the United States. Key advances such as the release of the first draft of the human genome,[3,4] the use of microarray to identify large deletions or duplications of DNA,[5] and the development of ES to identify pathogenic variants at the nucleotide level[6,7] serve as milestones in the history of genomic medicine. These advances have driven the current era of Mendelian disease diagnostics in which genetic testing can define genetically heterogeneous syndromes that are indistinguishable by clinical findings alone and provide a precise diagnosis. Making a precise genetic diagnosis can support disease-targeted surveillance or therapy,[8–10] facilitate more accurate genetic counseling about natural history and recurrence risks for a larger number of conditions,[11,12] and allow individuals to take part in precision or N-of-1 therapies.[13]

As new technologies have enabled greater access to inexpensive sequencing-based genetic tests, best practice guidelines have been modified to encourage appropriate use and emphasize their strengths. For example, demonstration that the use of chromosomal microarray to detect copy number variants (CNVs) had clinical utility beyond the characterization of cancer[14] led to guidelines supporting its use to evaluate individuals with suspected genetic syndromes.[15] Subsequently, in 2010, the American College of Medical Genetics (ACMG) modified these guidelines and recommended the use of microarray as a first-tier test for individuals with developmental delay or congenital anomalies[16] and, in 2021, again changed these guidelines to reflect the value of using ES as a first- or second-tier test to make a precise genetic diagnosis in persons with congenital anomalies, developmental delay, or intellectual disability.[17]

Current approaches to identify a precise molecular diagnosis in an individual suspected to have a genetic condition might include CNV analysis by microarray to identify large deletions or duplications, and/or a phenotype-informed gene panel or ES **(Table 1)**. However, 50–60% of individuals with a suspected Mendelian condition remain undiagnosed after clinically available comprehensive genetic testing, including ES, although it should be noted that there is substantial variability in the diagnostic rate depending on the phenotype.[18–21] In the critical care setting, such as the neonatal or pediatric ICU, the diagnostic approach may be somewhat different: as time to a precise genetic diagnosis is often of greater utility, broad and comprehensive testing, such as ES, has shown clinical value, with rapid turnaround times favored.[22–24] Despite the use of rapid ES or genome sequencing (GS) early in the diagnostic evaluation, the diagnostic rates in cohorts of critically ill infants range from 20–60% depending on ascertainment criteria, with some stratification schemes resulting in higher diagnostic rates than others.[21,24–27]

Several factors contribute to the varied and overall modest diagnostic rate in individuals suspected of having a Mendelian condition who undergo clinical testing. First, the genetic basis of many Mendelian conditions remains unknown. Second, for conditions for which the underlying gene(s) is known, the test ordered might not interrogate the appropriate gene(s) (e.g., single-gene or multi-gene panels), variant type(s) (e.g., short tandem repeat expansion) or epigenetic signature(s) (e.g., methylation status). Third, technical limitations may make it difficult to identify a pathogenic variant (e.g., CNV detection from ES). Fourth, there may not be sufficient information to interpret the pathogenicity of a variant (e.g., variants of unknown significance [VUS]). Compounding the latter issue is that the interpretation of variants by diagnostic laboratories can vary substantially because of differences in how evidence of pathogenicity of a variant is weighted,[28] although standardization of classification and data sharing efforts should mitigate this effect.[29–31] Fifth, incomplete penetrance and challenges associated with distinguishing whether a phenotype is due to large-effect alleles or the result of complex inheritance patterns (e.g., digenic or oligogenic) makes it difficult to identify the molecular etiology of a phenotype. Finally, the diagnostic rate has historically been dependent on the depth of phenotypic information available at variant adjudication.[32] While overall guidelines for systematic phenotyping do not exist, proposals to use human phenotype ontology terms and phenopackets would provide a standard for phenotype sharing across laboratories, clinicians, and researchers.[33,34]

Limited options for genome-wide testing exist for individuals who remain without a precise genetic diagnosis after current clinical testing options have been exhausted. GS is increasingly available, and newer technologies such as RNA sequencing (RNA-seq), optical genome mapping (OGM), and long-read GS (lrGS) are emerging. However, the value of these technologies over current testing strategies has yet to be determined across multiple clinical contexts and settings. Nevertheless, alone or in various combinations, these technologies may offer advantages that complement or perhaps replace conventional genome-wide testing strategies. This begs the question, what is the next best step in evaluating an individual with a suspected Mendelian condition after negative clinical genetic testing? Herein, we discuss testing options when a precise genetic diagnosis cannot be made via conventional testing, provide examples of how emerging technologies could be used to make a precise genetic diagnosis, and provide guidance to clinicians about the use of these technologies.

**Exome sequencing and reanalysis**

The use of clinical ES has substantially increased diagnostic rates across a broad range of categorical phenotypes and for Mendelian conditions in general, ranging from 25–40% depending on the phenotype and setting,[20,35,36] with higher diagnostic rates reported in populations in which consanguinity is common.[37] Thus, while the diagnostic yield is relatively high, on aggregate, more than 50% of individuals tested remain undiagnosed after clinical ES.[38–40] Reanalysis of existing ES data may uncover a pathogenic variant years after the data were generated **(Table 2)** due to new gene discovery for Mendelian conditions, resolution of VUS as pathogenic, and improvements in bioinformatic variant calling pipelines.

Yields from reanalysis of ES data vary widely depending on the age of the data, with those generated 5–10 years ago having a higher yield owing to the number of novel disease genes described in the intervening time period and the types of analyses applied.[41] A recent systematic review identified an increased diagnostic yield of approximately 15% across 27 studies and recommended reanalysis 18 months after the original analysis to optimize yield.[42] Diagnoses identified via ES reanalysis can be divided into two broad categories: variants missed by analysis pipelines and variants that were previously identified but not considered diagnostic. The first category often includes variants for which current variant-calling pipelines have limited technical sensitivity and/or reliability, such as indels, noncoding variants in regions flanking the coding segments targeted by the exome sequencing capture kit,[43] or CNVs.[41,44]

Many of the diagnoses made by reanalysis of ES data involve reinterpretation of previously detected variants with new evidence supporting their pathogenicity.[41,44–48] Such variants may be in a gene whose function was unknown at the time of original analysis or that had limited evidence to support the link between the gene and the condition. Criteria for reporting variants in genes not currently associated with Mendelian disease can vary between clinical laboratories, and therefore, these diagnoses are often found during research reanalysis.[49] However, diagnoses found on exome reanalysis may also be in known disease genes not previously thought to explain the phenotype, where the clinical interpretation of a variant has changed due to novel data such as additional clinical information, new variant inheritance information, segregation data from other affected family members, newly published cases, or an expansion of the phenotype associated with a gene.[46] In this regard, clinician input is often critical to making the diagnosis[50] and can also lead to detection of a second or additional genetic diagnosis, especially in cases with clinical findings not fully explained by a single Mendelian condition.[41,51]

Exome reanalysis is now formally recommended by the ACMG[52] and may be requested by the treating clinician, undertaken in the research setting,[49] or conducted by a clinical laboratory at regular intervals.[53] The ACMG recommends that clinical laboratories prioritize reanalysis "to maximize the potential clinical impact," such as for variants initially classified as a VUS and for reevaluating variants when relevant resources become newly available (e.g., population control genetic databases, published gene–disease relationships, or variant assessment methods).[52] However, specific policies regarding reanalysis, including frequency and communicating results to clinicians and patients and their families, are left to the discretion of individual laboratories. If exome reanalysis is not revealing, then pooling of exome-negative cases with similar phenotypes can be used in gene discovery efforts.

Several limitations of ES should be considered when deciding whether an individual would benefit from reanalysis or if other testing should be considered instead **(Table 3)**. Reanalysis of ES data may not identify mosaic variants underlying a condition as they are difficult to detect without deeper coverage or sequencing an affected tissue. For example, some lymphatic malformations are due to variants that have allele frequencies less than 1%.[54] In addition, known pathogenic variants, such as intronic or promoter variants which fall outside the protein-coding regions, may be missed on ES but identified by gene panels that are deliberately designed to capture regions containing these variants.

**Short-read genome sequencing**
Compared to ES, srGS provides a more unbiased sampling of the entire genome, providing a platform that can identify a spectrum of clinically relevant variation and enable the analysis of coding and noncoding variants. This allows for the detection of coding variants in regions poorly covered by ES and improves detection of SVs including CNVs, copy-neutral events such as inversions, and short tandem repeats (STRs). While srGS is increasingly being offered on a clinical basis, the interpretation of variants beyond those that can be identified by ES is limited. This is due to challenges in predicting the pathogenicity of noncoding variants, such as those suspected to affect splicing or gene expression that are less often reported in variant databases.[55] Similarly, interpreting the pathogenicity of SVs is challenging due to limited data on their population frequencies[56] and on the predicted phenotype from novel SVs that impact multiple genes. Thus, despite the ability to capture a wider spectrum of pathogenic variation, varying analytical and reporting practices by clinical labs temper the current added utility of srGS. Although best practices for variant reporting have been developed[57], identification of noncoding variants as well as SVs by lrGS may lead to greater discrepancy in reporting across clinical laboratories than is seen with ES.

Incremental diagnostic yields for srGS vary across studies and can depend on factors such as the characteristics of the cohort selected, age and quality of prior sequencing, unique aspects of the phenotype being studied, or the analytical tools used to call variants. Several studies have demonstrated a modest increase in diagnostic yield (5–20%) for srGS when performed after nondiagnostic ES **(Table 2)**.[58–60] For example, in a cohort of individuals with Alagille syndrome and previously nondiagnostic ES, srGS successfully identified pathogenic variants, including SVs and a noncoding SNV.[61,62] We anticipate that adoption of comprehensive variant calling pipelines by clinical laboratories combined with expanded variant databases, especially for noncoding variation, is likely to improve diagnostic utility.

Limitations of srGS include increased data generation costs, higher analytical burden, and a lower likelihood of identifying mosaic variants when compared to ES because of lower average coverage. In addition, the availability of clinical srGS is more limited due to payor restrictions.

Through competition between sequencing manufacturers and improvements in informatics, data processing, and sequencing chemistries, srGS is becoming increasingly cost effective and may soon be more cost effective than ES due to the ability to detect multiple types of variants with a single test.[63–65] As with clinical ES, we anticipate that as more studies show improved diagnostic yield and simplified testing, guidelines may shift toward recommending srGS as a first-tier test, with the majority of payors following closely behind.

**Targeted and whole-genome long-read sequencing**

Long-read sequencing (LRS) technology produces individual DNA or RNA reads ranging from 1 kb to several megabases in length.[66] There are two commercial long-read sequencing technologies currently available, one offered by Pacific Biosciences (PacBio) and the other by Oxford Nanopore Technologies (ONT).[67] PacBio sequencing works by monitoring a polymerase as it replicates a circular piece of DNA.[68] While the technology is error-prone at the base level, high-quality reads can be produced by combining multiple read segments from the same DNA molecule into a single, high-quality consensus read, which limits the average read length to approximately 15 kbp. ONT sequencing works by measuring changes in current as a single-stranded DNA or RNA molecule passes through a protein nanopore. This produces reads with a higher per-base error rate than PacBio HiFi reads, but they can be significantly longer[67,69] and more rapidly analyzed because signal may be decoded while sequencing.[70,71] Detection of CpG methylation is possible using both technologies when original DNA molecules are sequenced with no additional modifications during library preparation.[72,73] In comparison to short reads, long reads map better to repetitive regions of the genome, simplify identification of pathogenic SVs such as repeat expansions, and allow for phasing of variants.[74,75]

To reduce costs and simplify analysis, targeted LRS (T-LRS) of high-priority regions using either a Cas9-based approach on both platforms[76,77] or Adaptive Sampling on the ONT platform[78–80] has been shown to be effective in identifying missing variants in specific genes of interest **(Table 2)**. The benefit of T-LRS in clinical testing has become less certain as the cost of whole-genome LRS (lrGS) falls. This is because preparing samples for targeted sequencing can require additional time, such as for the design of guide RNAs for Cas9-based approaches, shearing of DNA for Adaptive Sampling, or the need for multiple sequencing runs if coverage is insufficient.

T-LRS requires identification of a candidate region prior to use. In cases in which no target region has been identified, lrGS represents an agnostic approach to identifying novel genes or loci of interest, but with higher cost and increased computational and interpretation burden above even srGS. Despite the challenges with data management, storage, and analysis, both T-LRS and lrGS are expected to have advantages in calling SNVs, indels, and SVs over ES or srGS.[66,74] This is because it is easier to reliably map long reads to complex regions of the genome and to then call variants within these regions; however, variant callers for lrGS are less mature than callers for ES/srGS.[75] LRS allows more SV breakpoints to be estimated with higher resolution and better identification of clinically relevant repeat expansions.[60,79,81] Several studies have shown that improvements in variant detection when using long-read technology can facilitate identification of genotype–phenotype associations in genomic regions that could not be analyzed using short reads.[60,82,83] Limited population-level data exist for SVs, SNVs, and indels in regions refractory to analysis using short reads. Ongoing efforts such as those from the *All of Us* project and lrGS sequencing of samples from the 1000 Genomes Project will address this limitation but will take several years to complete. In addition, as with srGS data, there are few tools for interpretation of noncoding SNVs, indels, and SVs.

Additional clinically relevant advantages of LRS over short-read sequencing exist. For example, both T-LRS and lrGS have been used for phasing of *de novo* variants or when parental samples were not available.[60,79] Because original DNA molecules are often sequenced, LRS data can be used to simultaneously evaluate both sequence and methylation status using a single data source (**Table 2**).[72,84] Unfortunately, similar to challenges with the large number of SVs identified by LRS, variation associated with methylation status in the population is unknown, leading to a need for databases containing tissue, age, and gender-matched controls for filtering and analysis.

When to use LRS to evaluate a challenging clinical case remains unclear. Frequently changing pipelines and limited reference datasets (especially from diverse populations) for filtering and prioritizing of variants identified by LRS creates challenges.[82,85–87] Further benchmarking efforts are also needed to identify sequencing artifacts and to standardize tools before widespread clinical application. The eventual adoption of LRS for clinical use will depend on curation of variants identified by LRS in control populations and side-by-side comparisons of the incremental diagnostic yield of srGS compared to lrGS in nondiagnostic cases. While studies support increased variant detection by lrGS in individuals without a precise genetic diagnosis via ES[60,79] and also in medically relevant genes,[83] further work is needed to assess the clinical utility of the technology compared to existing methods.

**Optical genome mapping**

Optical genome mapping (OGM) is a technique that works by imaging fluorescent labels that have been enzymatically introduced at canonical sequences on long, megabase-sized, DNA molecules.[88,89] The pattern of fluorescent labels are compared between the sample and a reference genome for identification of SVs. This allows for the detection of SVs that are challenging to detect with other methods, such as CNVs smaller than 25 kb and balanced events like inversions or translocations (**Table 2**).[90] A direct comparison of lrGS, srGS, and OGM on the same sample showed that 1 in 3 deletions and 3 of 4 insertions larger than 10 kb were detectable only by OGM,[82] a result that should be revisited with newer lrGS datasets and SV calling pipelines.

An early attempt at demonstrating clinical utility of OGM showed 100% concordance with previously detected SVs in Duchenne muscular dystrophy (OMIM: 310200), including a 5.1-Mb inversion (which had previously required combined PCR, MLPA, RNA-Seq and srGS to decipher), as well as determination of carrier status in maternal samples.[91] Several case reports have further highlighted the usefulness of OGM in identifying SVs difficult to detect with other technologies[92–96] and in karyotyping.[93] Successful application of OGM to resolve haplotypes and size in the 3.3-kb repeat arrays causative of facioscapulohumeral muscular dystrophy (FSHD, OMIM: 158900)[97,98] has led to the first CLIA-approved application of OGM (PerkinElmer/Bionano Genomics EnFocus FSHD Analysis), with a goal of replacing traditional Southern blots for clinical testing.

While OGM excels as a single technology to detect SVs, large CNVs, and complex rearrangements, there are limitations. Sequence information is not available and resolution is limited by the spacing of fluorescent tags along the genome and by the resolution of the imaging photocell. Extraction of high-molecular-weight DNA is required for optimal results, similar to some LRS-based approaches. For clinical use, detecting variants with OGM is only the first step. As seen with srGS and lrGS, determining the clinical relevance of SVs remains a major challenge.

**RNA sequencing**

Though srGS and lrGS can capture a wide variety of variants, interpreting the impact of many intronic and noncoding variants can be challenging despite the development of advanced algorithms such as SpliceAI[99] or Genomiser,[100] leading to potentially pathogenic variants being missed. Thus, RNA sequencing may be used to identify the gene responsible for a disease based on expression or splicing without prior knowledge of underlying variants (**Table 2**).[101–104] Combined with DNA variant knowledge, this information can be used to clarify the impact of a candidate splice or other noncoding variant, to identify a missing allele in a known recessive disease, or to identify candidate genes in individuals with nondiagnostic prior testing. RNA sequencing can also identify other types of clinically relevant variation, including variants that affect RNA stability, differences in polyadenylation, novel transcripts, or variants in non-protein-coding genes that may not be evaluated by standard analysis pipelines.[105]

Transcriptome profiles, however, vary greatly depending on the tissue sampled and clinical status of the affected individual. Not all genes or isoforms are expressed in easily accessible tissues, such as blood or fibroblasts, and thus may not be interrogatable by RNA sequencing. Despite these limitations, RNA sequencing has been used successfully to reclassify VUSs and identify missing pathogenic variants with a diagnostic rate reported between 7.5% to 34% depending on the phenotype studied and the tissue sequenced.[101,106–109] In general, work has shown that more genes are expressed in fibroblasts than from WBCs, however work has shown that these are less relevant for immunological phenotypes[109,110] and some neurologic phenotypes, where lymphoblastoid cell lines were sufficient.[111] In certain phenotypes, the overall diagnostic rate of RNA sequencing may be similar to or better than ES. One recent study compared the diagnostic rate of these two methods in a cohort of individuals with neuromuscular disease and found that RNA sequencing alone of muscle tissue identified a higher number of pathogenic variants (38.1%, 24/86) than ES alone (34.9%, 22/96).[112]

Other challenges with widespread implementation of RNA sequencing exist beyond tissue specificity of expression and isoform usage. First, detecting differences in expression that may point to potentially causative genes requires careful selection of controls because of both biological variation (e.g., age and environment) as well as variation in experimental protocols and sequencing platforms,[113] which has led some to recommend the development of in-house control sets to address the experimental component.[108] Second, allele-specific expression (ASE) at a specific locus or across an entire chromosome, such as in cases with skewed X-chromosome inactivation, has been proposed as an underlying cause of phenotypic variability or disease severity in rare disease and can be identified with RNA sequencing.[114–116] Unfortunately, predicting ASE from variant-level DNA sequencing data alone remains challenging[113] and best practices for outlier detection are emerging, thus ASE detection currently relies on inclusion of both RNA sequencing and variant-level DNA data to detect transcripts that detect one variant at a higher level than another.[117] Third, gene fusion events have been identified in previously unsolved cases, but identification requires additional analysis steps that are not usually undertaken.[118]

While not yet clinically available, long-read RNA sequencing of amplicons, cDNA or original RNA molecules may simplify isoform analysis and permit easier identification of fusion transcripts.[119–122] For example, one recent study used LRS of amplicons from an individual with Werner syndrome to determine exactly which haplotype was affected by exon skipping, allowing the group to more closely evaluate the other haplotype for pathogenic variants.[80] Although cDNA sequencing is available on both the PacBio and Nanopore platforms, direct RNA sequencing is currently available only on the Nanopore platform.[123,124] This method can simultaneously assay expression, isoform structure, and RNA modifications, which opens new possibilities in evaluating individuals who remain undiagnosed after extensive evaluation. Several research consortia are exploring long-read RNA sequencing after nondiagnostic ES, srGS, and short-read RNA sequencing, thus we anticipate ongoing advances in this space over time.

**DNA Methylation Analysis**

Multiple Mendelian conditions are caused by dysregulation of the epigenetic machinery, with subsequent alteration of DNA methylation patterns.[125] These conditions are associated with distinct DNA methylation signatures, or 'episignatures,' which can be used to distinguish between different syndromes.[126] Episignatures can thus be used to support variant reclassification or to suggest a specific Mendelian condition in individuals with previous nondiagnostic testing (**Table 2**).[127] For example, Aref-Eshgi and colleagues[128] applied genome-wide DNA methylation analysis to develop a computational model to support the diagnosis of fourteen neurodevelopmental conditions with known episignatures, including Coffin-Siris syndrome and other BAFopathies, CHARGE syndrome, and Kabuki syndrome. Using this model, they were able to resolve 21 (31%) of 67 individuals who presented with ambiguous clinical features and/or genetic findings suspicious for one of these Mendelian conditions, including individuals with no candidate variants found on molecular testing.

While such studies highlight the potential of DNA methylation analysis to resolve undiagnosed cases, there are important limitations. At present, DNA methylation testing is commonly done using peripheral blood samples, and available reference datasets used to identify episignatures are also generally derived from blood samples. As epigenetic profiles can vary substantially between different tissues,[129] results observed from blood samples may be different with respect to their generalizability for the disease-related tissue of interest. Moreover, there is the need for consensus analytical standards for DNA methylation testing, including correcting for age, sex, environmental influences, and other factors that can impact results of DNA methylation analysis.[130] In addition, existing DNA methylation array technologies have been noted to have limited ability for detecting low-grade genetic mosaicism (<20%)[127,131] and can only identify Mendelian conditions with previously characterized episignatures, both of which can reduce diagnostic sensitivity. Nonetheless, DNA methylation testing has been shown to resolve as many as 30% of individuals with features suspicious for rare neurodevelopmental conditions, a yield comparable to the solve rates reported for chromosomal microarray (15–20%) and ES (30–40%).[132] This suggests that it may be worth incorporating methylation profiling as part of the first-line diagnostic workup of individuals with specific phenotypes, such as neurodevelopmental disorders, suspected imprinting disorders, repeat expansion disorders, or a VUS in a known methylation gene. Because LRS can concurrently identify methylation status when generating sequence data, we expect episignatures to be incorporated into these analysis pipelines in the future.

**Integrating biochemical and proteomic data: Multiomics approaches**

Using a combination of data types, such as genomic, transcriptomic, epigenetic, proteomic, metabolomic, or biochemical data, is referred to as a multiomic approach.[133,134] While integrating the large amounts of data that can be generated with any of these approaches seems challenging, the overall idea is straightforward: look for overlapping clues in the data available for variant identification and prioritization. Many consortia studying challenging unsolved cases have used this approach successfully to guide reanalysis efforts. For example, within the Undiagnosed Diseases Network (UDN) an individual was suspected to have deficiency of 3-hydroxy-3-methylglutaryl coenzyme A lyase (encoded by *HMGL*) based on persistently high urine organic acid levels.[19] Reanalysis of exome sequencing data revealed a deletion in the first exon of *HMGL* and RNA sequencing confirmed that *HMGL* expression was 50% that of unaffected controls. Other groups have shown that enzymatic or metabolomic assays have utility in the interpretation of VUSs identified by panel testing.[135]

A multiomics approach can be especially useful in cases with suspected mitochondrial disease given the phenotypic heterogeneity that can be observed.[136,137] Mitochondrial disease can arise from pathogenic changes in genes in either the nuclear DNA or the 16.5-kb mitochondrial genome (mtDNA). While pathogenic variants in the nuclear genome are detectable by regular genomic sequencing platforms such as ES, identifying pathogenic variants in mtDNA may require careful evaluation because of tissue-specific heteroplasmy. Pathogenic variants in mtDNA may directly impact protein-coding genes (e.g., Leber hereditary optic neuropathy) or tRNA genes (e.g., MELAS: mitochondrial encephalopathy, myopathy, lactic acidosis, and stroke-like episodes) or cause large genomic rearrangements of mtDNA (e.g., Kearns-Sayre syndrome). Testing options for mtDNA-specific variants range from conventional technologies such as targeted sequencing, Southern blotting or array comparative genomic hybridization to high-coverage short-read sequencing to interrogate the entire mtDNA for SNVs/indels and SVs.[138] Analysis of mtDNA variations may be combined with biochemical results, such as electron transport chain assays, metabolic profiling, or proteomics, to achieve accurate variant interpretation (**Table 2**).

Both targeted and global metabolomic data have been used to either aid in the prioritization of variants identified by ES or to suggest specific genes or pathways for evaluation, although the yield of the latter has been low.[139,140] For example, in a retrospective study of 170 patients, untargeted metabolomics contributed toward prioritization of variants from ES in 74 individuals (43.5%) and confirmed clinical diagnosis in 21 cases, yielding a diagnostic rate of 12.3%.[140] Several software packages have been developed to aid in the integration of metabolic data with existing data types.[141–143] A major limitation of untargeted metabolomic data is the challenge of finding appropriate controls as differences in age, diet, or medication usage can alter clinically relevant metabolomic profiles.

Finally, proteomic analysis can provide valuable insights into the genes or pathways that may be affected in an individual with a suspected Mendelian condition.[144] This is especially true in cases where the affected cell or tissue can be easily collected. For example, Grabowski and colleagues performed proteome analysis of individuals with monogenic diseases affecting neutrophil function.[145] Surprisingly, large proteome changes were observed in only some, but not all, known conditions and observed changes did not correlate with transcriptome analysis, demonstrating the power of orthogonal data in elucidating changes in rare conditions. Overall, the integration of different omics technologies that complement one another can provide key clues in individuals that remain undiagnosed after extensive clinical testing. The specific test ordered should be driven by the patient's phenotype and candidate variants.

**DISCUSSION**

Advances in genetic testing provide opportunity and hope for individuals who remain undiagnosed after comprehensive clinical testing. However, appropriate application of these technologies, which may only be available in the research setting, remains unclear. Thus, our aim is to both provide an overview of each of these new technologies **(Table 1)** and to provide a list of options about what next steps in testing exist for individuals who lack a precise genetic diagnosis after ES **(Figure 1, Table 4)**. Although certain technologies are not yet clinically available, collaboration between clinicians and researchers is essential for rare disease diagnosis, and familiarity with these emerging techniques may facilitate both referral to an appropriate research study or clinical implementation once a new technology is available.

Careful reevaluation of prior genetic and laboratory testing of the individual with a suspected Mendelian condition may itself be high yield. This includes ensuring any prior VUSs are not now explanatory and that any candidate genes have not recently been associated with a similar phenotype. Prior candidate variants or genes should be shared via Matchmaker Exchange[146] to facilitate identification of similar cases that can strengthen associations or phenotype expansions. Exome reanalysis, if possible, should be undertaken at least once, especially if it has been more than one year since the initial test or last reanalysis. Easily overlooked tests, such as karyotype or microarray, should be considered if prior testing may have missed variants that could be identified using these modalities.

Determining the next best step depends on several factors and should be considered on a case-by-case basis. In <u>cases with a candidate gene</u>, such as when a single variant has been identified in a gene associated with a recessive Mendelian condition or no variants were identified in a case with strong biochemical or phenotypic evidence pointing to a single gene or small number of genes associated with a dominant Mendelian condition, then evaluation for a missing variant should be undertaken **(Figure 1, Table 4)**. Clinical testing options include targeted RNA-sequencing, methylation analysis if the suspected Mendelian condition is associated with a distinct epigenetic signature, or srGS. Prior to ordering srGS, the provider should confirm whether the testing laboratory will analyze and report variants that are beyond what would typically be reported by ES, such as deep intronic variants, regulatory variants, and SVs. If clinical options are not available, T-LRS (ONT) or lrGS (PacBio or ONT) is potentially the next-best test to be performed on a research basis as it offers simultaneous evaluation of coding/noncoding SNVs, indels, repeat expansions/constrictions and SVs that may be missed by srGS as well as providing variant phasing and methylation changes.[79,80] We anticipate that future studies will provide data to better guide the decision-making process in these cases.

For individuals <u>without a clear candidate gene or variant</u> to explore, a broad approach should be taken **(Figure 1, Table 4)**. In these cases, clinicians should consider methylation analysis, srGS, or RNA-sequencing, while keeping in mind that empirical data as to which test has the highest diagnostic yield in this setting is limited. Clinical suspicion and test availability can be a guide, such as ordering methylation analysis for individuals in which a Mendelian condition with a distinct epigenetic signature is suspected. Similarly, srGS may be a better choice than RNA sequencing for individuals with isolated neurologic phenotypes, since the variant responsible might be in a gene expressed only in the brain and would therefore be difficult to identify using RNA sequencing of readily available tissues (e.g., fibroblasts). Furthermore, only a limited number of clinical laboratories offer untargeted RNA sequencing at this time and no systematic evaluation of their results has been undertaken. Given the unproven clinical utility, most payors do not currently reimburse for these tests and institutional policies may dictate whether they can be ordered. Because of these limitations, it may be best to refer the individual to a research program focused on families without a precise genetic diagnosis **(Table 5)**.

The options in the decision tree presented here highlight the complexity of evaluating individuals with suspected Mendelian conditions who lack a precise genetic diagnosis **(Figure 1)**. Current testing approaches require multiple steps, which may involve repeated clinical visits and require individuals and their caregivers to take time off work, travel long distances, and be subjected to multiple studies and tests. Costs associated with travel and/or time away from work may result in individuals delaying or deferring visits and testing, resulting in a system that is not equitable and that provides no clear benefit to any one participant.[147] This leads to a question of whether and how new technologies can be used to simplify the clinical testing process by reducing the number of individual tests required, with the goal of reducing the time to diagnosis and increasing the diagnostic rate. Additionally, dual diagnoses where more than one genetic diagnosis is identified in an individual have been reported in up to 5% of patients and are often challenging to diagnose because of the presentation of a blended phenotype.[148,149] In the near future, we anticipate that a single test, such as srGS, will be used to simplify the evaluation process and reduce inequities in care, with lrGS replacing or supplementing this data as costs fall over time.[150]

Carefully designed studies will be needed to determine if one next-best test exists after negative ES or if the choice of what test to pursue is best determined by underlying phenotype or clinical suspicion. These studies will be supported by the collection and biobanking of biospecimens from those individuals affected by rare Mendelian conditions as both resources for benchmarking as well as for understanding novel mechanisms or genes that underpin these unsolved cases. The development of new reference genomes that permit telomere-to-telomere analysis of patient genomes will need to be considered and likely lead to novel gene–phenotype associations in previously inaccessible genomic regions.[151,152] Over time, the current standard diagnostic evaluation pathway will likely change, with a focus on simplifying overall testing and evaluation of previously "challenging" regions or variants. Thus, we envision a time when a single data source, such as srGS or lrGS, is evaluated in a stepwise fashion, perhaps enhanced by concurrent methylome or transcriptome, or metabolomic analysis that replaces the time-consuming progression of microarray, panel testing, and ES. Both patients and providers may then benefit from simplified testing with decreased time to diagnosis and, ideally, greater understanding of the molecular underpinnings of rare diseases.

# FIGURES

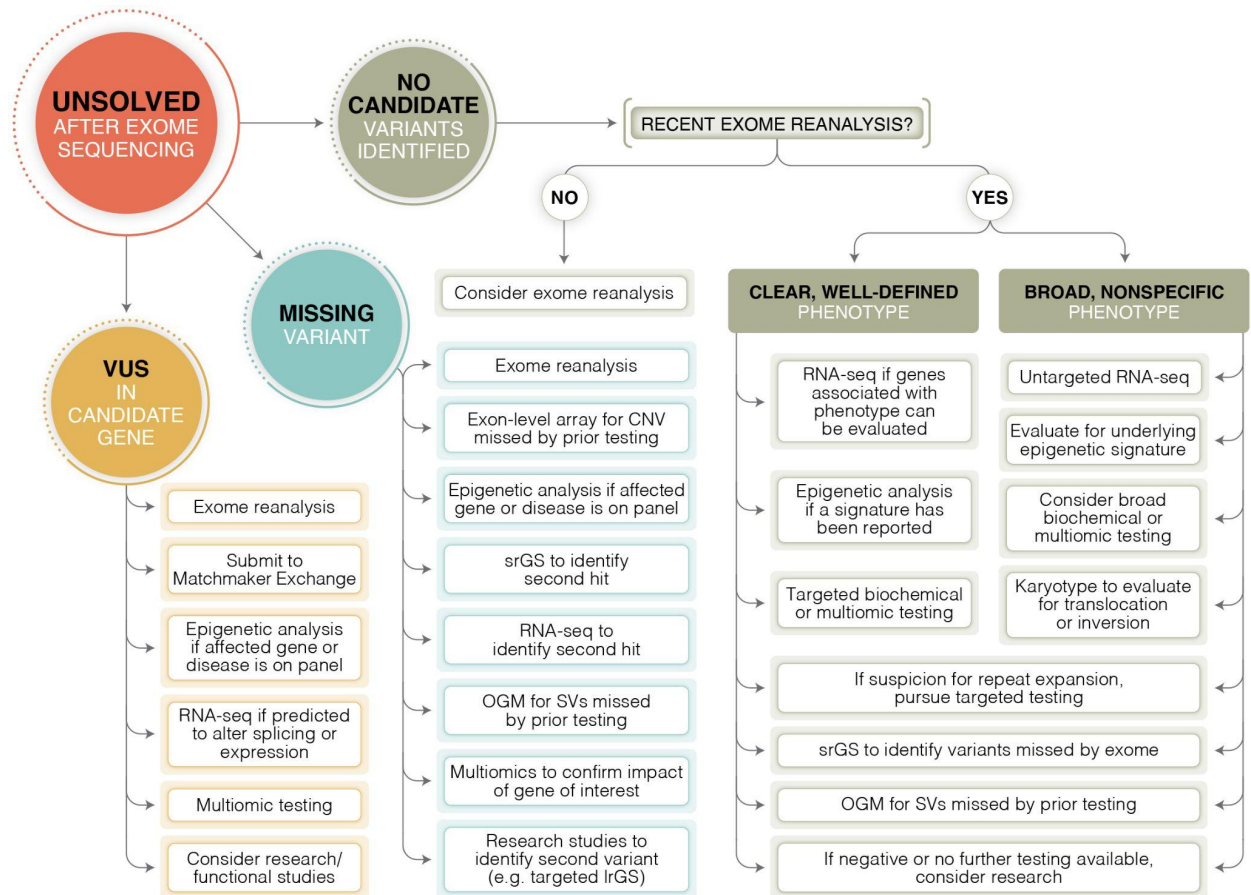

**Figure 1**
Testing paths and options for individuals with clinical findings that cannot be partially or fully explained by a precise genetic diagnosis after exome sequencing. In each path, exome reanalysis should be considered first. Many options are similar among the various paths but are of highest diagnostic yield at different steps of the evaluation process.

# TABLES

## Table 1. Advantages, limitations, and use cases of different types of genetic testing technologies.

CNV (copy number variant); SV (structural variant); ES (exome sequencing); LRS (long-read sequencing); OGM (optical genome mapping); turnaround time (TAT); srGS (short-read genome sequencing); lrGS (long-read genome sequencing); CDG (congenital disorder of glycosylation); IBD (identity by descent); NIPT (noninvasive prenatal testing).

| | Advantages | Limitations | Examples of when to consider use |
|---|---|---|---|
| **Commonly used in clinical genetics** | **Karyotype** | | |
| | • Inexpensive<br>• Rapid TAT<br>• Detection of mosaic events<br>• Detection of balanced rearrangements difficult to detect by sequencing methods | • Difficult to identify modest or small SVs such as deletions, duplications, or inversions<br>• Cannot detect SNVs or indels | • Recurrent pregnancy loss<br>• When a translocation is suspected (e.g., multiple deletions/duplications found on microarray)<br>• To resolve discrepancies between prenatal testing (e.g., NIPT) and phenotype (e.g., by U/S) or atypical genitalia at birth<br>• Suspected aneuploidy or mosaicism (e.g., compare affected and unaffected skin) |
| | **Microarray** | | |
| | • Inexpensive<br>• Rapid TAT<br>• Identify regions of homozygosity or IBD | • May not identify small deletions or duplications<br>• Unable to identify inversions or translocations<br>• Multiple assays available – not all tests are equivalent | • Developmental delay or intellectual disability<br>• Multiple congenital anomalies<br>• Suspected sex chromosome aneuploidy |
| | **Gene panel or targeted gene sequencing** | | |
| | • Likely to have fewer incidental findings<br>• May capture regulatory or intronic regions not covered by ES<br>• May cover difficult-to-sequence or complex regions not covered by ES<br>• May include repeat expansions using orthogonal technologies<br>• Reporting criteria may return more VUSs in target genes | • Often singleton only, thus may require additional steps to confirm inheritance<br>• May not include genes recently associated with phenotype(s)<br>• CNV/SV analysis may depend on lab offering the test<br>• Gene panels offered vary depending on lab offering test | • High suspicion for a specific phenotype or Mendelian condition<br>• After negative ES if specific non-coding or intronic variants are suspected<br>• Suspicion for a Mendelian condition not well captured by short read technology |
| | **ES** | | |
| | • Broad evaluation of most protein-coding genes<br>• Data can undergo reanalysis<br>• May include CNV/SV analysis<br>• May detect mosaic variants | • May not include regulatory or intronic regions that are known to be associated with a specific phenotype<br>• Sensitivity for CNVs including one or few exons is poor<br>• Variant reporting based on lab's understanding of provided phenotype | • Negative prior testing<br>• Phenotype that does not fit a well-described syndrome<br>• Need for broad and rapid evaluation (e.g. in the intensive care unit)<br>• Panel testing unavailable for phenotype |
| | | • Multiple types of exomes are available, meaning not all tests are equivalent | |
| **Emerging use in clinical genetics** | **Methylation analysis** | | |
| | • Targeted or genome-wide evaluation<br>• May detect mosaic changes associated with phenotype<br>• Additional conditions can be evaluated on a research basis | • Limited number of conditions can be evaluated clinically<br>• Cannot identify causative DNA variant<br>• Signature may vary by condition and be affected by age or acute illness | • Clarification of VUS identified by prior testing<br>• Individuals with DD or ID after non-diagnostic ES<br>• Multiple congenital anomalies after non-diagnostic panel or ES<br>• Clinical suspicion of epigenetic disorder with negative prior evaluation |
| | **OGM** | | |
| | • High sensitivity for CNVs, SVs, and rearrangements | • Inability to detect SNVs, indels | • Suspicion for complex chromosomal event missed by prior testing |
| | **srGS** | | |
| | • Sampling entire genome<br>• Able to identify protein-coding, intronic, and regulatory variants<br>• Higher sensitivity for small CNV/SV identification than ES<br>• Able to identify SV breakpoints | • Intronic and regulatory regions may not be analyzed or interpreted depending on the lab performing the test<br>• Lower depth of coverage limits detection of mosaic variants and variants in difficult to sequence regions<br>• Higher likelihood of incidental and uncertain findings | • After negative ES<br>• As alternative to ES when need for broad and rapid evaluation<br>• Emerging as initial test in suspected Mendelian conditions |
| | **RNA sequencing** | | |
| | • Can be used to adjudicate VUSs that are predicted to affect splicing or expression<br>• Identify changes that impact splicing or expression but that are difficult to predict computationally<br>• Identify changes specific to tissue of interest<br>• Identify global changes in gene expression or splicing | • Require large number of controls<br>• A limited number of labs offer this clinically<br>• Tissue of interest may not be clinically accessible<br>• Methods for interpretation and reporting not standardized | • DNA candidate variant suggests impact on splicing or expression<br>• Identify variant in second allele in a recessive disease where genomic sequencing returned only one pathogenic variant<br>• Suspicion of phenotype with global effect on splicing or expression |
| **Research use** | **LRS (targeted or whole-genome)** | | |
| | • Unbiased sampling of entire genome<br>• SV identification and interpretation is better than srGS<br>• DNA methylation information is generated concomitant with sequence information<br>• Characterization of repetitive regions of the genome without targeted capture or computational tools | • Potentially more expensive than other modalities<br>• Analysis may require substantial compute and network resources<br>• Lack of publicly available population-level data for filtering and interpretation | • After non-diagnostic ES/GS and/or RNA seq<br>• Missing variant cases<br>• Phenotype suggests repeat expansion mechanism |

**Table 2. Examples of cases solved by specific technologies after prior testing was negative.**

| Category | Example |
|---|---|
| Diagnosis made by reanalysis of exome sequencing data | A three-year-old boy was noted to have unique facial features, developmental delay, anxiety, gastrointestinal dysmotility, and poor growth. While a Mendelian condition was suspected, clinical ES at age four was nondiagnostic. Reanalysis of the ES data identified a pathogenic variant in *PPM1D*, consistent with a diagnosis of Jansen de Vries syndrome. This disease–gene association had been made two years after the clinical exome was sent and one year prior to the reanalysis.[49] |
| srGS detects variants missed by ES | Shortly after birth, a neonate was noted to have features consistent with early-onset Marfan syndrome (OMIM: 154700). Sequencing and deletion-duplication analysis (via multiplex ligation-dependent probe amplification, MLPA) of *FBN1* did not reveal any pathogenic variants. Subsequent trio exome sequencing with CNV analysis was similarly nondiagnostic. srGS identified a heterozygous 385-base pair deletion in *FBN1* involving the early-onset Marfan syndrome critical region (exons 24-32). This small SV was not identified on CNV analysis of ES data due to quality filtering.[62] |
| Targeted LRS (T-LRS) reveals a deletion not identified by clinical testing | An individual suspected to have Hermansky-Pudlak syndrome (OMIM 203300) based on clinical features had a negative clinical microarray followed by trio exome sequencing that identified a single paternally inherited pathogenic variant in *HPS1*, the gene associated with this recessive disease. T-LRS on the ONT platform identified the paternally inherited pathogenic variant as well as a 1,900-bp frameshifting deletion not identified by either microarray or exome.[79] This deletion was clinically validated using an exon-level array. |
| Simultaneous evaluation of repeat length and methylation with LRS | In an individual with Baratela-Scott syndrome (OMIM: 615777) known to have an expansion in the promoter of *XYLT1* that leads to silencing of the gene, T-LRS simultaneously detected expansion of the repeat and associated CpG hypermethylation in the proband as well as low-level silencing of the premutation allele in the mother.[79] |
| Simultaneous identification of a deletion and inversion with OGM | In a young child with epileptic encephalopathy that remained undiagnosed after chromosome microarray, an epilepsy panel that included an exon-level array for *CDKL5*, ES, and srGS, Cope and colleagues used OGM to identify a mosaic deletion and inversion in *CDKL5* and to estimate that the deletion and inversion were present in approximately 25% of DNA molecules assayed.[94] The result was clinically confirmed with short-read mate-pair sequencing. |
| RNA sequencing detects a pathogenic splice variant not identified by ES | Hong and colleagues used RNA sequencing to evaluate a cohort of individuals with neuromuscular disease and nondiagnostic clinical testing.[112] In an individual with recurrent rhabdomyolysis and nondiagnostic ES, RNA-seq detected an exon-skipping event in *LPIN1*. The causative variant was found to be a synonymous variant in the last exon of the gene that was not predicted to be splice-altering by computational tools. This highlights the challenge with interpreting rare synonymous variants whose impact is not predicted using standard analysis tools. |
| Using epigenetic signatures to diagnose rare Mendelian conditions | In a study of 207 individuals referred for clinical genome-wide DNA methylation testing, epigenetic signatures were used to associate 57 cases with one of 50 previously known conditions. The majority of individuals (48/57) carried a VUS in a gene associated with the disease represented by the epigenetic signature.[127] |
| Multiomics approach aids variant interpretation | Deletions and duplications have been reported to cause lethal perinatal mitochondrial disease at the *ATAD3* locus, but they are difficult to analyze given the repetitive nature of the region. In individuals with suspected mitochondrial disease, Frazier and colleagues used a combination of ES, srGS, lrGS, and quantitative proteomics to evaluate 17 individuals from 16 families and identified six different *de novo* duplications in the *ATAD3* locus associated with the phenotype in these individuals.[153] |

**Table 3. Considerations and next steps after nondiagnostic clinical exome sequencing.**
ES (exome sequencing); CNV (copy number variant); SV (structural variant); OGM (optical genome mapping); srGS (short-read genome sequencing); lrGS (long-read genome sequencing)

| Type of variant | Action that can be taken |
| --- | --- |
| Small CNV not detectable by ES | Exon-level array may identify small CNVs. Alternatively, srGS or lrGS may detect small CNVs. |
| Regulatory variant located in a region not captured by ES | Depending on the specificity of the phenotype, consider more-targeted gene testing that includes sequencing of regulatory regions or srGS; consider RNA-seq or epigenetic signature testing. |
| Deep intronic variant that affects splicing | Consider a panel that may include known intronic variants. Either srGS, lrGS, or RNA-seq can also be used to identify or confirm the variant. |
| Variant in a gene not previously associated with the phenotype, not assessed and/or reported because of laboratory analysis and/or reporting criteria | Consider submission to Matchmaker exchange[146] and referral to research group or consortia who can conduct a broader, gene discovery-oriented analysis (e.g., flag putatively deleterious variants in genes not previously associated with human disease, and identify additional cases who harbor variants in this candidate gene) |
| Structural variant not detected by ES (e.g., a complex rearrangement or inversion) | OGM, srGS, or lrGS can be used to identify and clarify SVs missed by exome sequencing. |
| Repeat expansion | Depending on the specificity of the phenotype, consider a disease-targeted panel/gene testing or srGS or lrGS with repeat expansion detection. |
| Variants that cannot be phased | If parental samples not available or a variant is *de novo* then clinical srGS may phase if variants are close enough together. If not, either mate-pair sequencing or lrGS can be used. |
| Mosaic variants | Discuss reporting criteria and technical thresholds for variant calling with laboratory. If there is strong clinical suspicion for a specific genetic disease, consider targeted testing with higher sequencing depth. Consider high-depth exome sequencing; multiple-tissue/multiple-sample sequencing. |

**Table 4. Clinical and research options for specific classes of variants.**
OGM (optical genome mapping); srGS (short-read whole-genome sequencing); LRS (long-read sequencing); lrGS (long-read whole-genome sequencing); SV (structural variant).

| Variant Type | Clinical Testing Options | Research Testing Options |
|---|---|---|
| **Missing Variant**<br><br>(one variant in AR condition, no variants in AD or XLD condition with clear phenotype) | • **Reanalysis of existing data:** may identify new variants or new gene–phenotype associations<br>• **Microarray or Exon-level array:** may identify CNVs missed by prior testing<br>• **Targeted sequencing with del/dup:** panels may include regions not analyzed by prior testing such as intronic or regulatory regions<br>• **RNA sequencing:** evaluate for splicing or expression difference that could identify a missing variant or confirm the suspected diagnosis<br>• **Methylation analysis:** could be used to confirm suspected condition if there is an associated difference in methylation<br>• **OGM:** may identify SVs missed by prior testing<br>• **srGS:** may capture intronic variants, regulatory variants, or SVs not identified by prior testing | • **Reanalysis of existing data:** reanalysis on a research basis may identify additional variants or be used in gene discovery efforts<br>• **srGS:** may identify additional variants when performed on a research basis<br>• **RNA sequencing:** evaluation of splicing or expression differences could identify or confirm a missing variant<br>• **LRS:** Targeted or lrGS may identify intronic or regulatory variants, a missed SV, or differences in methylation that could affect function of the gene |
| **Variants that Cannot be Phased**<br><br>(*de novo* variants or parental samples not available) | • **Single gene testing:** if variants are close together, they may be captured and phased by short-read sequencing; discuss with lab prior to sending | • **Mate-pair sequencing:** may allow for phasing of variants<br>• **LRS:** reads from targeted or lrGS may be long enough to phase or allow for phasing using other nearby variants |
| **Structural Variant**<br><br>(to identify exact breakpoints or additional variants) | • **Karyotype:** if additional large-scale variants are suspected such as translocations, large inversions, or large CNVs<br>• **Microarray or Exon-level array:** evaluate for additional CNVs if not already done<br>• **srGS:** may be used to identify exact breakpoints of an SV or additional variants if lab is able to report these<br>• **OGM:** could identify additional changes but would be unlikely to provide more precise breakpoints | • **Mate-pair sequencing:** may be used to identify additional SVs or confirm breakpoints of known SVs<br>• **LRS:** Targeted or lrGS can be used to identify precise breakpoints and additional SVs as well as methylation changes associated with the SV |
| **Variant in Candidate Gene**<br><br>(gene is not clearly associated with the phenotype) | • **Additional testing:** consider other tests that may identify variants in genes previously associated with the phenotype but missed by prior testing<br>• **Methylation analysis:** may identify methylation pattern similar to other well-described condition and help with interpretation of candidate gene | • **Functional studies:** may confirm pathogenicity and that the variant gives a similar phenotype as the affected individual<br>• **Matchmaker exchange:** sharing candidate gene may identify other individuals with similar phenotype and variants in the same gene |

| Variant of Uncertain Significance (pathogenicity of the variant is not established) | <ul><li>**Reanalysis of existing data:** variant may have been reported in separate case since initially identified and could be upgraded</li><li>**RNA sequencing:** could be used to upgrade variant in cases where the variant is suspected to affect splicing or expression</li><li>**Methylation analysis:** could be used to confirm pathogenicity if the variant is in a gene with an associated difference in methylation</li><li>**Biochemical testing:** assessment of specific biomarkers to confirm an uncertain diagnosis</li></ul> | <ul><li>**Functional studies:** may confirm pathogenicity and that the variant gives a similar phenotype as the affected individual</li><li>**Matchmaker exchange:** sharing candidate gene may identify other individuals with similar phenotype and variants in the same gene</li></ul> |
| --- | --- | --- |

**Table 5. A subset of global diagnostic programs for individuals with rare or unsolved genetic diseases.** Not all programs remain active. Table adapted from Cloney *et al*.[154]

| Program | Country/Region | Website |
|---|---|---|
| Rare and Undiagnosed Diseases Diagnostic Service[158] | Australia | https://www.australiangenomics.org.au/research/the-australian-undiagnosed-diseases-network/ |
| Care4Rare Canada[155] | Canada | https://www.care4rare.ca/ |
| RD-Connect[160] | Europe | https://rd-connect.eu/ |
| Solve-RD | Europe | https://solve-rd.eu/ |
| National Center for Rare diseases (NCRD)[162] | Italy | https://www.udnpitaly.com/pagine-13-about_us |
| The Initiative on Rare and Undiagnosed Diseases (IRUD)[159] | Japan | https://www.amed.go.jp/en/program/IRUD/ |
| Korean Undiagnosed Diseases Program (KUDP)[157] | Korea | n/a |
| Rare Disease Genomics in South Africa | South Africa | https://www.sun.ac.za/english/faculty/healthsciences/Molecular_Biology_Human_Genetics/rare-disease_genomics/research-projects |
| SpainUDP[161] | Spain | https://spainudp.isciii.es/home/ |
| Deciphering Developmental Disorders[156] | United Kingdom | https://www.ddduk.org/ |
| GREGoR Consortium | United States | https://gregorconsortium.org/ |
| Undiagnosed Diseases Program[19] | United States | https://undiagnosed.hms.harvard.edu/ |

**Data Availability Statement**
No primary data was generated for this manuscript.

**Acknowledgements**
We thank Angela Miller for assistance with figure preparation.

**Funding**
DEM is supported by NIH grant DP5OD033357. MHW is supported by NIH/NICHD K23HD102589 and an Early Career Award from the Thrasher Research Fund. The GREGoR Consortium is funded by the National Human Genome Research Institute of the National Institutes of Health, through the following grants: U01HG011758, U01HG011755, U01HG011745, U01HG011762, U01HG011744, and U24HG011746. The content is solely the responsibility of the authors and does not necessarily represent the official views of the National Institutes of Health.

**Author Contributions**
Conceptualization: M.H.W., C.M.R., S.M., M.M., M.H.D., H.B., B.Y., E.E.G., E.C.D., D.J., A.S-J., L.S., M.T., S.B.M, M.T.W., S.I.B., A.O-L., F.J.S., D.E.M. Writing-original draft: M.H.W., C.M.R., S.M., M.M., M.H.D., H.B., B.Y., E.E.G., E.C.D., D.J., A.S-J., L.S., M.T., S.B.M, M.T.W., S.I.B., A.O-L., F.J.S., D.E.M. Writing-review & editing: all authors.

**Ethics Declaration**
No human subjects, live vertebrates, or higher invertebrate research was undertaken as part of this manuscript.

**Conflict of Interest**
C.M.R. is a consultant for My Gene Counsel. H.B. is a shareholder of Bionano Genomics Inc, Pacific Biosciences Inc and Illumina Inc. B.Y. has received royalties from UpToDate. A.O-L. is on the SAB of Congenica. S.B.M. is a consultant for BioMarin, MyOme and Tenaya Therapeutics. M.E.T. consults for BrigeBio Pharma and receives research funding and/or reagents from Illumina Inc., Levo Therapeutics, and Microsoft Inc. M.T.W. holds stock in Personalis, Inc. F.J.S. has received travel support to speak on behalf of ONT and PacBio. D.E.M. holds stock options in MyOme and is engaged in a research agreement with ONT and they have paid for him to travel to speak on their behalf. M.H.W, S.M., M.M., M.H.D, P.B., E.E.G, E.C.D, D.J., A.S-J., L.M.S., M.J.B., J.X.C., and S.I.B. declare no conflicts.

# REFERENCES

1.  Karczewski KJ, Francioli LC, Tiao G, et al. The mutational constraint spectrum quantified from variation in 141,456 humans. *Nature*. 2020;581(7809):434-443.

2.  Landrum MJ, Lee JM, Benson M, et al. ClinVar: improving access to variant interpretations and supporting evidence. *Nucleic Acids Res*. 2018;46(D1):D1062-D1067.

3.  Lander ES, Linton LM, Birren B, et al. Initial sequencing and analysis of the human genome. *Nature*. 2001;409(6822):860-921.

4.  Venter JC, Adams MD, Myers EW, et al. The sequence of the human genome. *Science*. 2001;291(5507):1304-1351.

5.  Oostlander AE, Meijer GA, Ylstra B. Microarray-based comparative genomic hybridization and its applications in human genetics. *Clin Genet*. 2004;66(6):488-495.

6.  Ng SB, Bigham AW, Buckingham KJ, et al. Exome sequencing identifies MLL2 mutations as a cause of Kabuki syndrome. *Nat Genet*. 2010;42(9):790-793.

7.  Bamshad MJ, Ng SB, Bigham AW, et al. Exome sequencing as a tool for Mendelian disease gene discovery. *Nat Rev Genet*. 2011;12(11):745-755.

8.  Al-Khatib SM, Stevenson WG, Ackerman MJ, et al. 2017 AHA/ACC/HRS guideline for management of patients with ventricular arrhythmias and the prevention of sudden cardiac death: Executive summary: A Report of the American College of Cardiology/American Heart Association Task Force on Clinical Practice Guidelines and the Heart Rhythm Society. *Heart Rhythm*. 2018;15(10):e190-e252.

9.  Mazzanti A, Maragna R, Vacanti G, et al. Interplay Between Genetic Substrate, QTc Duration, and Arrhythmia Risk in Patients With Long QT Syndrome. *J Am Coll Cardiol*. 2018;71(15):1663-1671.

10. Clark MM, Stark Z, Farnaes L, et al. Meta-analysis of the diagnostic and clinical utility of genome and exome sequencing and chromosomal microarray in children with suspected genetic diseases. *NPJ Genom Med*. 2018;3:16.

11. Reuter CM, Kohler JN, Bonner D, et al. Yield of whole exome sequencing in undiagnosed patients facing insurance coverage barriers to genetic testing. *J Genet Couns*. 2019;28(6):1107-1118.

12. Zastrow DB, Zornio PA, Dries A, et al. Exome sequencing identifies de novo pathogenic variants in FBN1 and TRPS1 in a patient with a complex connective tissue phenotype. *Cold Spring Harb Mol Case Stud*. 2017;3(1):a001388.

13. Kim J, Hu C, Moufawad El Achkar C, et al. Patient-Customized Oligonucleotide Therapy for a Rare Genetic Disease. *N Engl J Med*. 2019;381(17):1644-1652.

14. Shaffer LG, Beaudet AL, Brothman AR, et al. Microarray analysis for constitutional cytogenetic abnormalities. *Genet Med*. 2007;9(9):654-662.

15. Manning M, Hudgins L, Professional Practice and Guidelines Committee. Array-based technology and recommendations for utilization in medical genetics practice for detection of chromosomal abnormalities. *Genet Med*. 2010;12(11):742-745.

16. Miller DT, Adam MP, Aradhya S, et al. Consensus statement: chromosomal microarray is a first-tier clinical diagnostic test for individuals with developmental disabilities or congenital anomalies. *Am J Hum Genet*. 2010;86(5):749-764.

17. Manickam K, McClain MR, Demmer LA, et al. Exome and genome sequencing for pediatric patients with congenital anomalies or intellectual disability: an evidence-based clinical guideline of the American College of Medical Genetics and Genomics (ACMG). *Genet Med*. 2021;23(11):2029-2037.

18. Shashi V, McConkie-Rosell A, Rosell B, et al. The utility of the traditional medical genetics diagnostic evaluation in the context of next-generation sequencing for undiagnosed genetic disorders. *Genet Med*. 2014;16(2):176-182.

19. Splinter K, Adams DR, Bacino CA, et al. Effect of Genetic Diagnosis on Patients with Previously Undiagnosed Disease. *N Engl J Med*. 2018;379(22):2131-2139.

20. Yang Y, Muzny DM, Xia F, et al. Molecular findings among patients referred for clinical whole-exome sequencing. *JAMA*. 2014;312(18):1870-1879.

21. Meng L, Pammi M, Saronwala A, et al. Use of Exome Sequencing for Infants in Intensive Care Units: Ascertainment of Severe Single-Gene Disorders and Effect on Medical Management. *JAMA Pediatr*. 2017;171(12):e173438.

22. The NICUSeq Study Group, Krantz ID, Medne L, et al. Effect of Whole-Genome Sequencing on the Clinical Management of Acutely Ill Infants With Suspected Genetic Disease: A Randomized Clinical Trial. *JAMA Pediatr*. 2021;175(12):1218-1226.

23. Farnaes L, Hildreth A, Sweeney NM, et al. Rapid whole-genome sequencing decreases infant morbidity and cost of hospitalization. *npj Genomic Medicine*. 2018;3(1):1-8.

24. Kingsmore SF, Cakici JA, Clark MM, et al. A Randomized, Controlled Trial of the Analytic and Diagnostic Performance of Singleton and Trio, Rapid Genome and Exome Sequencing in Ill Infants. *Am J Hum Genet*. 2019;105(4):719-733.

25. Gubbels CS, VanNoy GE, Madden JA, et al. Prospective, phenotype-driven selection of critically ill neonates for rapid exome sequencing is associated with high diagnostic yield. *Genet Med*. 2020;22(4):736-744.

26. Australian Genomics Health Alliance Acute Care Flagship, Lunke S, Eggers S, et al. Feasibility of Ultra-Rapid Exome Sequencing in Critically Ill Infants and Children With Suspected Monogenic Conditions in the Australian Public Health Care System. *JAMA*. 2020;323(24):2503-2511.

27. Maron JL, Kingsmore SF, Wigby K, et al. Novel Variant Findings and Challenges Associated With the Clinical Integration of Genomic Testing: An Interim Report of the Genomic Medicine for Ill Neonates and Infants (GEMINI) Study. *JAMA Pediatr*. 2021;175(5):e205906.

28. O'Daniel JM, McLaughlin HM, Amendola LM, et al. A survey of current practices for genomic sequencing test interpretation and reporting processes in US laboratories. *Genet Med*. 2017;19(5):575-582.

29. Harrison SM, Dolinsky JS, Knight Johnson AE, et al. Clinical laboratories collaborate to resolve differences in variant interpretations submitted to ClinVar. *Genet Med*. 2017;19(10):1096-1104.

30. Balmaña J, Digiovanni L, Gaddam P, et al. Conflicting Interpretation of Genetic Variants and Cancer Risk by Commercial Laboratories as Assessed by the Prospective Registry of Multiplex Testing. *J Clin Oncol*. 2016;34(34):4071-4078.

31. Rehm HL, Berg JS, Brooks LD, et al. ClinGen--the Clinical Genome Resource. *N Engl J Med*. 2015;372(23):2235-2242.

32. Johnson B, Ouyang K, Frank L, et al. Systematic use of phenotype evidence in clinical genetic testing reduces the frequency of variants of uncertain significance. *Am J Med Genet A*. 2022;188(9):2642-2651.

33. Köhler S, Gargano M, Matentzoglu N, et al. The Human Phenotype Ontology in 2021. *Nucleic Acids Res*. 2021;49(D1):D1207-D1217.

34. Jacobsen JOB, Baudis M, Baynam GS, et al. The GA4GH Phenopacket schema defines a computable representation of clinical data. *Nat Biotechnol*. 2022;40(6):817-820.

35. Yang Y, Muzny DM, Reid JG, et al. Clinical whole-exome sequencing for the diagnosis of mendelian disorders. *N Engl J Med*. 2013;369(16):1502-1511.

36. Retterer K, Juusola J, Cho MT, et al. Clinical application of whole-exome sequencing across clinical indications. *Genet Med*. 2016;18(7):696-704.

37. Monies D, Abouelhoda M, Assoum M, et al. Lessons Learned from Large-Scale, First-Tier Clinical Exome Sequencing in a Highly Consanguineous Population. *Am J Hum Genet*. 2019;104(6):1182-1201.

38. Mainali A, Athey T, Bahl S, et al. Diagnostic yield of clinical exome sequencing in adulthood in medical genetics clinics. *Am J Med Genet A*. Published online November 19, 2022. doi:10.1002/ajmg.a.63053

39. Seo GH, Kim T, Choi IH, et al. Diagnostic yield and clinical utility of whole exome sequencing using an automated variant prioritization system, EVIDENCE. *Clin Genet*. 2020;98(6):562-570.

40. Lee H, Deignan JL, Dorrani N, et al. Clinical exome sequencing for genetic identification of rare Mendelian disorders. *JAMA*. 2014;312(18):1880-1887.

41. Liu P, Meng L, Normand EA, et al. Reanalysis of Clinical Exome Sequencing Data. *N Engl J Med*. 2019;380(25):2478-2480.

42. Tan NB, Stapleton R, Stark Z, et al. Evaluating systematic reanalysis of clinical genomic data in rare disease from single center experience and literature review. *Mol Genet Genomic Med*. 2020;8(11):e1508.

43. Romero R, de la Fuente L, Del Pozo-Valero M, et al. An evaluation of pipelines for DNA variant detection can guide a reanalysis protocol to increase the diagnostic ratio of genetic diseases. *NPJ Genom Med*. 2022;7(1):7.

44. Fung JLF, Yu MHC, Huang S, et al. A three-year follow-up study evaluating clinical utility of exome sequencing and diagnostic potential of reanalysis. *NPJ Genom Med*. 2020;5(1):37.

45. Berger SI, Miller I, Tochen L. Recessive GCH1 Deficiency Causing DOPA-Responsive Dystonia Diagnosed by Reported Negative Exome. *Pediatrics*. 2022;149(2). doi:10.1542/peds.2021-052886

46. Al-Nabhani M, Al-Rashdi S, Al-Murshedi F, et al. Reanalysis of exome sequencing data of intellectual disability samples: Yields and benefits. *Clin Genet*. 2018;94(6):495-501.

47. Al-Murshedi F, Meftah D, Scott P. Underdiagnoses resulting from variant misinterpretation: Time for systematic reanalysis of whole exome data? *Eur J Med Genet*. 2019;62(1):39-43.

48. Gordeeva V, Sharova E, Babalyan K, Sultanov R, Govorun VM, Arapidi G. Benchmarking germline CNV calling tools from exome sequencing data. *Sci Rep*. 2021;11(1):14416.

49. Schmitz-Abe K, Li Q, Rosen SM, et al. Unique bioinformatic approach and comprehensive reanalysis improve diagnostic yield of clinical exomes. *Eur J Hum Genet*. 2019;27(9):1398-1405.

50. Basel-Salmon L, Orenstein N, Markus-Bustani K, et al. Improved diagnostics by exome sequencing following raw data reevaluation by clinical geneticists involved in the medical care of the individuals tested. *Genet Med*. 2019;21(6):1443-1451.

51. Karaca E, Posey JE, Coban Akdemir Z, et al. Phenotypic expansion illuminates multilocus pathogenic variation. *Genet Med*. 2018;20(12):1528-1537.

52. Deignan JL, Chung WK, Kearney HM, et al. Points to consider in the reevaluation and reanalysis of genomic test results: a statement of the American College of Medical Genetics and Genomics (ACMG). *Genet Med*. 2019;21(6):1267-1270.

53. Leung ML, Ji J, Baker S, et al. A Framework of Critical Considerations in Clinical Exome Reanalyses by Clinical and Laboratory Standards Institute. *J Mol Diagn*. 2022;24(2):177-188.

54. Zenner K, Jensen DM, Dmyterko V, et al. Somatic activating BRAF variants cause isolated lymphatic malformations. *HGG Adv*. 2022;3(2):100101.

55. Costain G, Walker S, Marano M, et al. Genome Sequencing as a Diagnostic Test in Children With Unexplained Medical Complexity. *JAMA Netw Open*. 2020;3(9):e2018109.

56. Collins RL, Brand H, Karczewski KJ, et al. A structural variation reference for medical and population genetics. *Nature*. 2020;581(7809):444-451.

57. Austin-Tse CA, Jobanputra V, Perry DL, et al. Best practices for the interpretation and reporting of clinical whole genome sequencing. *NPJ Genom Med*. 2022;7(1):27.

58. Alfares A, Aloraini T, Subaie LA, et al. Whole-genome sequencing offers additional but limited clinical utility compared with reanalysis of whole-exome sequencing. *Genet Med*.

2018;20(11):1328-1333.

59. Lionel AC, Costain G, Monfared N, et al. Improved diagnostic yield compared with targeted gene sequencing panels suggests a role for whole-genome sequencing as a first-tier genetic test. *Genet Med*. 2018;20(4):435-443.

60. Cohen ASA, Farrow EG, Abdelmoity AT, et al. Genomic answers for children: Dynamic analyses of >1000 pediatric rare disease genomes. *Genet Med*. Published online March 15, 2022. doi:10.1016/j.gim.2022.02.007

61. Rajagopalan R, Gilbert MA, McEldrew DA, et al. Genome sequencing increases diagnostic yield in clinically diagnosed Alagille syndrome patients with previously negative test results. *Genet Med*. 2021;23(2):323-330.

62. Wojcik MH, Thiele K, Grant CF, et al. Genome Sequencing Identifies the Pathogenic Variant Missed by Prior Testing in an Infant with Marfan Syndrome. *J Pediatr*. 2019;213:235-240.

63. Lavelle TA, Feng X, Keisler M, et al. Cost-effectiveness of exome and genome sequencing for children with rare and undiagnosed conditions. *Genet Med*. 2022;24(6):1349-1361.

64. Incerti D, Xu XM, Chou JW, Gonzaludo N, Belmont JW, Schroeder BE. Cost-effectiveness of genome sequencing for diagnosing patients with undiagnosed rare genetic diseases. *Genet Med*. 2022;24(1):109-118.

65. Almogy G, Pratt M, Oberstrass F, et al. Cost-efficient whole genome-sequencing using novel mostly natural sequencing-by-synthesis chemistry and open fluidics platform. *bioRxiv*. Published online August 10, 2022:2022.05.29.493900. doi:10.1101/2022.05.29.493900

66. Coster WD, De Coster W, Weissensteiner MH, Sedlazeck FJ. Towards population-scale long-read sequencing. *Nature Reviews Genetics*. 2021;22(9):572-587. doi:10.1038/s41576-021-00367-3

67. Logsdon GA, Vollger MR, Eichler EE. Long-read human genome sequencing and its applications. *Nat Rev Genet*. 2020;21(10):597-614.

68. Wenger AM, Peluso P, Rowell WJ, et al. Accurate circular consensus long-read sequencing improves variant detection and assembly of a human genome. *Nat Biotechnol*. 2019;37(10):1155-1162.

69. Clarke J, Wu HC, Jayasinghe L, Patel A, Reid S, Bayley H. Continuous base identification for single-molecule nanopore DNA sequencing. *Nat Nanotechnol*. 2009;4(4):265-270.

70. Galey M, Reed P, Wenger T, et al. 3-hour genome sequencing and targeted analysis to rapidly assess genetic risk. *medRxiv*. Published online September 13, 2022:2022.09.09.22279746. doi:10.1101/2022.09.09.22279746

71. Gorzynski JE, Goenka SD, Shafin K, et al. Ultrarapid Nanopore Genome Sequencing in a Critical Care Setting. *N Engl J Med*. 2022;386(7):700-702.

72. Simpson JT, Workman RE, Zuzarte PC, David M, Dursi LJ, Timp W. Detecting DNA cytosine methylation using nanopore sequencing. *Nat Methods*. 2017;14(4):407-410.

73. Flusberg BA, Webster DR, Lee JH, et al. Direct detection of DNA methylation during single-molecule, real-time sequencing. *Nat Methods*. 2010;7(6):461-465.

74. Mahmoud M, Gobet N, Cruz-Dávalos DI, Mounier N, Dessimoz C, Sedlazeck FJ. Structural variant calling: the long and the short of it. *Genome Biology*. 2019;20(1). doi:10.1186/s13059-019-1828-7

75. Sedlazeck FJ, Lee H, Darby CA, Schatz MC. Piercing the dark matter: bioinformatics of long-range sequencing and mapping. *Nat Rev Genet*. 2018;19(6):329-346.

76. Gilpatrick T, Lee I, Graham JE, et al. Targeted nanopore sequencing with Cas9-guided adapter ligation. *Nat Biotechnol*. 2020;38(4):433-438.

77. Walsh T, Casadei S, Munson KM, et al. CRISPR-Cas9/long-read sequencing approach to identify cryptic mutations in BRCA1 and other tumour suppressor genes. *J Med Genet*. 2021;58(12):850-852.

78. Stevanovski I, Chintalaphani SR, Gamaarachchi H, et al. Comprehensive genetic diagnosis of tandem repeat expansion disorders with programmable targeted nanopore sequencing. *Sci Adv*. 2022;8(9):eabm5386.

79. Miller DE, Sulovari A, Wang T, et al. Targeted long-read sequencing identifies missing disease-causing variation. *Am J Hum Genet*. 2021;108(8):1436-1449.

80. Miller DE, Lee L, Galey M, et al. Targeted long-read sequencing identifies missing pathogenic variants in unsolved Werner syndrome cases. *J Med Genet*. Published online May 9, 2022. doi:10.1136/jmedgenet-2022-108485

81. Cretu Stancu M, van Roosmalen MJ, Renkens I, et al. Mapping and phasing of structural variation in patient genomes using nanopore sequencing. *Nat Commun*. 2017;8(1):1326.

82. Chaisson MJP, Sanders AD, Zhao X, et al. Multi-platform discovery of haplotype-resolved structural variation in human genomes. *Nat Commun*. 2019;10(1):1784.

83. Wagner J, Olson ND, Harris L, et al. Curated variation benchmarks for challenging medically relevant autosomal genes. *Nat Biotechnol*. 2022;40(5):672-680.

84. Lee I, Razaghi R, Gilpatrick T, et al. Simultaneous profiling of chromatin accessibility and methylation on human cell lines with nanopore sequencing. *Nat Methods*. 2020;17(12):1191-1199.

85. Beyter D, Ingimundardottir H, Oddsson A, et al. Long-read sequencing of 3,622 Icelanders provides insight into the role of structural variants in human diseases and other traits. *Nat Genet*. 2021;53(6):779-786.

86. Ebert P, Audano PA, Zhu Q, et al. Haplotype-resolved diverse human genomes and integrated analysis of structural variation. *Science*. 2021;372(6537). doi:10.1126/science.abf7117

87. Audano PA, Sulovari A, Graves-Lindsay TA, et al. Characterizing the Major Structural Variant Alleles of the Human Genome. *Cell*. 2019;176(3):663-675.e19.

88. Bocklandt S, Hastie A, Cao H. Bionano Genome Mapping: High-Throughput, Ultra-Long

Molecule Genome Analysis System for Precision Genome Assembly and Haploid-Resolved Structural Variation Discovery. In: Suzuki Y, ed. *Single Molecule and Single Cell Sequencing*. Springer Singapore; 2019:97-118.

89. Lam ET, Hastie A, Lin C, et al. Genome mapping on nanochannel arrays for structural variation analysis and sequence assembly. *Nat Biotechnol*. 2012;30(8):771-776.

90. Chan S, Lam E, Saghbini M, et al. Structural Variation Detection and Analysis Using Bionano Optical Mapping. *Methods Mol Biol*. 2018;1833:193-203.

91. Barseghyan H, Tang W, Wang RT, et al. Next-generation mapping: a novel approach for detection of pathogenic structural variants with a potential utility in clinical diagnosis. *Genome Med*. 2017;9(1):90.

92. Schnause AC, Komlosi K, Herr B, et al. Marfan Syndrome Caused by Disruption of the FBN1 Gene due to A Reciprocal Chromosome Translocation. *Genes* . 2021;12(11). doi:10.3390/genes12111836

93. Sabatella M, Mantere T, Waanders E, et al. Optical genome mapping identifies a germline retrotransposon insertion in SMARCB1 in two siblings with atypical teratoid rhabdoid tumors. *J Pathol*. 2021;255(2):202-211.

94. Cope H, Barseghyan H, Bhattacharya S, et al. Detection of a mosaic CDKL5 deletion and inversion by optical genome mapping ends an exhaustive diagnostic odyssey. *Mol Genet Genomic Med*. 2021;9(7):e1665.

95. Chen M, Zhang M, Qian Y, et al. Identification of a likely pathogenic structural variation in the LAMA1 gene by Bionano optical mapping. *NPJ Genom Med*. 2020;5:31.

96. Fahiminiya S, Rivard GE, Scott P, et al. A full molecular picture of F8 intron 1 inversion created with optical genome mapping. *Haemophilia*. 2021;27(5):e638-e640.

97. Stence AA, Thomason JG, Pruessner JA, et al. Validation of Optical Genome Mapping for the Molecular Diagnosis of Facioscapulohumeral Muscular Dystrophy. *J Mol Diagn*. 2021;23(11):1506-1514.

98. Dai Y, Li P, Wang Z, et al. Single-molecule optical mapping enables quantitative measurement of D4Z4 repeats in facioscapulohumeral muscular dystrophy (FSHD). *J Med Genet*. 2020;57(2):109-120.

99. Jaganathan K, Kyriazopoulou Panagiotopoulou S, McRae JF, et al. Predicting Splicing from Primary Sequence with Deep Learning. *Cell*. 2019;176(3):535-548.e24.

100. Smedley D, Schubach M, Jacobsen JOB, et al. A Whole-Genome Analysis Framework for Effective Identification of Pathogenic Regulatory Variants in Mendelian Disease. *Am J Hum Genet*. 2016;99(3):595-606.

101. Frésard L, Smail C, Ferraro NM, et al. Identification of rare-disease genes using blood transcriptome sequencing and large control cohorts. *Nat Med*. 2019;25(6):911-919.

102. Ferraro NM, Strober BJ, Einson J, et al. Transcriptomic signatures across human tissues identify functional rare genetic variation. *Science*. 2020;369(6509). doi:10.1126/science.aaz5900

103. Pala M, Zappala Z, Marongiu M, et al. Population- and individual-specific regulatory variation in Sardinia. *Nat Genet*. 2017;49(5):700-707.

104. Lord J, Baralle D. Splicing in the Diagnosis of Rare Disease: Advances and Challenges. *Front Genet*. 2021;12:689892.

105. Montgomery SB, Bernstein JA, Wheeler MT. Toward transcriptomics as a primary tool for rare disease investigation. *Cold Spring Harb Mol Case Stud*. 2022;8(2). doi:10.1101/mcs.a006198

106. Cummings BB, Marshall JL, Tukiainen T, et al. Improving genetic diagnosis in Mendelian disease with transcriptome sequencing. *Sci Transl Med*. 2017;9(386). doi:10.1126/scitranslmed.aal5209

107. Lee H, Huang AY, Wang LK, et al. Diagnostic utility of transcriptome sequencing for rare Mendelian diseases. *Genet Med*. 2020;22(3):490-499.

108. Kremer LS, Bader DM, Mertes C, et al. Genetic diagnosis of Mendelian disorders via RNA sequencing. *Nat Commun*. 2017;8:15824.

109. Murdock DR, Dai H, Burrage LC, et al. Transcriptome-directed analysis for Mendelian disease diagnosis overcomes limitations of conventional genomic testing. *J Clin Invest*. 2021;131(1). doi:10.1172/JCI141500

110. Aicher JK, Jewell P, Vaquero-Garcia J, Barash Y, Bhoj EJ. Mapping RNA splicing variations in clinically accessible and nonaccessible tissues to facilitate Mendelian disease diagnosis using RNA-seq. *Genet Med*. 2020;22(7):1181-1190.

111. Rentas S, Rathi KS, Kaur M, et al. Diagnosing Cornelia de Lange syndrome and related neurodevelopmental disorders using RNA sequencing. *Genet Med*. 2020;22(5):927-936.

112. Hong SE, Kneissl J, Cho A, et al. Transcriptome-based variant calling and aberrant mRNA discovery enhance diagnostic efficiency for neuromuscular diseases. *J Med Genet*. Published online April 6, 2022. doi:10.1136/jmedgenet-2021-108307

113. Zhang Z, van Dijk F, de Klein N, et al. Feasibility of predicting allele specific expression from DNA sequencing using machine learning. *Sci Rep*. 2021;11(1):10606.

114. Gui B, Slone J, Huang T. Perspective: Is Random Monoallelic Expression a Contributor to Phenotypic Variability of Autosomal Dominant Disorders? *Front Genet*. 2017;8:191.

115. Sun Y, Luo Y, Qian Y, et al. Heterozygous Deletion of the SHOX Gene Enhancer in two Females With Clinical Heterogeneity Associating With Skewed XCI and Escaping XCI. *Front Genet*. 2019;10:1086.

116. Tukiainen T, Villani AC, Yen A, et al. Landscape of X chromosome inactivation across human tissues. *Nature*. 2017;550(7675):244-248.

117. Castel SE, Levy-Moonshine A, Mohammadi P, Banks E, Lappalainen T. Tools and best practices for data processing in allelic expression analysis. *Genome Biol*. 2015;16:195.

118. Oliver GR, Tang X, Schultz-Rogers LE, et al. A tailored approach to fusion transcript identification increases diagnosis of rare inherited disease. *PLoS One*.

2019;14(10):e0223337.

119. Byrne A, Beaudin AE, Olsen HE, et al. Nanopore long-read RNAseq reveals widespread transcriptional variation among the surface receptors of individual B cells. *Nat Commun*. 2017;8:16027.

120. Uapinyoying P, Goecks J, Knoblach SM, et al. A long-read RNA-seq approach to identify novel transcripts of very large genes. *Genome Res*. 2020;30(6):885-897.

121. De Roeck A, Van den Bossche T, van der Zee J, et al. Deleterious ABCA7 mutations and transcript rescue mechanisms in early onset Alzheimer's disease. *Acta Neuropathol*. 2017;134(3):475-487.

122. Nattestad M, Goodwin S, Ng K, et al. Complex rearrangements and oncogene amplifications revealed by long-read DNA and RNA sequencing of a breast cancer cell line. *Genome Res*. 2018;28(8):1126-1135.

123. Garalde DR, Snell EA, Jachimowicz D, et al. Highly parallel direct RNA sequencing on an array of nanopores. *Nat Methods*. 2018;15(3):201-206.

124. Jain M, Abu-Shumays R, Olsen HE, Akeson M. Advances in nanopore direct RNA sequencing. *Nat Methods*. 2022;19(10):1160-1164.

125. Fahrner JA, Bjornsson HT. Mendelian disorders of the epigenetic machinery: tipping the balance of chromatin states. *Annu Rev Genomics Hum Genet*. 2014;15:269-293.

126. Aref-Eshghi E, Rodenhiser DI, Schenkel LC, et al. Genomic DNA Methylation Signatures Enable Concurrent Diagnosis and Clinical Genetic Variant Classification in Neurodevelopmental Syndromes. *Am J Hum Genet*. 2018;102(1):156-174.

127. Sadikovic B, Levy MA, Kerkhof J, et al. Clinical epigenomics: genome-wide DNA methylation analysis for the diagnosis of Mendelian disorders. *Genet Med*. 2021;23(6):1065-1074.

128. Aref-Eshghi E, Bend EG, Colaiacovo S, et al. Diagnostic Utility of Genome-wide DNA Methylation Testing in Genetically Unsolved Individuals with Suspected Hereditary Conditions. *Am J Hum Genet*. 2019;104(4):685-700.

129. Laird PW. Principles and challenges of genomewide DNA methylation analysis. *Nat Rev Genet*. 2010;11(3):191-203.

130. Chater-Diehl E, Goodman SJ, Cytrynbaum C, Turinsky AL, Choufani S, Weksberg R. Anatomy of DNA methylation signatures: Emerging insights and applications. *Am J Hum Genet*. 2021;108(8):1359-1366.

131. Montano C, Britton JF, Harris JR, et al. Genome-wide DNA methylation profiling confirms a case of low-level mosaic Kabuki syndrome 1. *Am J Med Genet A*. 2022;188(7):2217-2225.

132. Srivastava S, Love-Nichols JA, Dies KA, et al. Meta-analysis and multidisciplinary consensus statement: exome sequencing is a first-tier clinical diagnostic test for individuals with neurodevelopmental disorders. *Genet Med*. 2019;21(11):2413-2421.

133. Stenton SL, Kremer LS, Kopajtich R, Ludwig C, Prokisch H. The diagnosis of inborn errors of metabolism by an integrative "multi-omics" approach: A perspective encompassing genomics, transcriptomics, and proteomics. *J Inherit Metab Dis*. 2020;43(1):25-35.

134. Khan S, Ince-Dunn G, Suomalainen A, Elo LL. Integrative omics approaches provide biological and clinical insights: examples from mitochondrial diseases. *J Clin Invest*. 2020;130(1):20-28.

135. Almeida LS, Pereira C, Aanicai R, et al. An integrated multiomic approach as an excellent tool for the diagnosis of metabolic diseases: our first 3720 patients. *Eur J Hum Genet*. Published online May 25, 2022. doi:10.1038/s41431-022-01119-5

136. Alston CL, Stenton SL, Hudson G, Prokisch H, Taylor RW. The genetics of mitochondrial disease: dissecting mitochondrial pathology using multi-omic pipelines. *J Pathol*. 2021;254(4):430-442.

137. Labory J, Fierville M, Ait-El-Mkadem S, Bannwarth S, Paquis-Flucklinger V, Bottini S. Multi-Omics Approaches to Improve Mitochondrial Disease Diagnosis: Challenges, Advances, and Perspectives. *Front Mol Biosci*. 2020;7:590842.

138. Zhang W, Cui H, Wong LJC. Comprehensive one-step molecular analyses of mitochondrial genome by massively parallel sequencing. *Clin Chem*. 2012;58(9):1322-1331.

139. Almontashiri NAM, Zha L, Young K, et al. Clinical Validation of Targeted and Untargeted Metabolomics Testing for Genetic Disorders: A 3 Year Comparative Study. *Sci Rep*. 2020;10(1):9382.

140. Alaimo JT, Glinton KE, Liu N, et al. Integrated analysis of metabolomic profiling and exome data supplements sequence variant interpretation, classification, and diagnosis. *Genet Med*. 2020;22(9):1560-1566.

141. Bongaerts M, Bonte R, Demirdas S, et al. Prioritizing disease-causing metabolic genes by integrating metabolomics with whole exome sequencing data. *bioRxiv*. Published online May 24, 2021. doi:10.1101/2021.05.21.21257573

142. Graham Linck EJ, Richmond PA, Tarailo-Graovac M, et al. metPropagate: network-guided propagation of metabolomic information for prioritization of metabolic disease genes. *NPJ Genom Med*. 2020;5:25.

143. Thistlethwaite LR, Li X, Burrage LC, et al. Clinical diagnosis of metabolic disorders using untargeted metabolomic profiling and disease-specific networks learned from profiling data. *Sci Rep*. 2022;12(1):6556.

144. Suhre K, McCarthy MI, Schwenk JM. Genetics meets proteomics: perspectives for large population-based studies. *Nat Rev Genet*. 2021;22(1):19-37.

145. Grabowski P, Hesse S, Hollizeck S, et al. Proteome Analysis of Human Neutrophil Granulocytes From Patients With Monogenic Disease Using Data-independent Acquisition. *Mol Cell Proteomics*. 2019;18(4):760-772.

146. Boycott KM, Azzariti DR, Hamosh A, Rehm HL. Seven years since the launch of the

Matchmaker Exchange: The evolution of genomic matchmaking. *Hum Mutat*. 2022;43(6):659-667.

147. Halley MC, Young JL, Fernandez L, et al. Perceived utility and disutility of genomic sequencing for pediatric patients: Perspectives from parents with diverse sociodemographic characteristics. *Am J Med Genet A*. 2022;188(4):1088-1101.

148. Posey JE, Rosenfeld JA, James RA, et al. Molecular diagnostic experience of whole-exome sequencing in adult patients. *Genet Med*. 2016;18(7):678-685.

149. Balci TB, Hartley T, Xi Y, et al. Debunking Occam's razor: Diagnosing multiple genetic diseases in families by whole-exome sequencing. *Clin Genet*. 2017;92(3):281-289.

150. Lowther C, Valkanas E, Giordano JL, et al. Systematic evaluation of genome sequencing as a first-tier diagnostic test for prenatal and pediatric disorders. *Cold Spring Harbor Laboratory*. Published online August 13, 2020:2020.08.12.248526. doi:10.1101/2020.08.12.248526

151. Nurk S, Koren S, Rhie A, et al. The complete sequence of a human genome. *Science*. 2022;376(6588):44-53.

152. Aganezov S, Yan SM, Soto DC, et al. A complete reference genome improves analysis of human genetic variation. *Science*. 2022;376(6588):eabl3533.

153. Frazier AE, Compton AG, Kishita Y, et al. Fatal perinatal mitochondrial cardiac failure caused by recurrent de novo duplications in the ATAD3 locus. *Med (N Y)*. 2021;2(1):49-73.

154. Cloney T, Gallacher L, Pais LS, et al. Lessons learnt from multifaceted diagnostic approaches to the first 150 families in Victoria's Undiagnosed Diseases Program. *J Med Genet*. 2022;59(8):748-758.

155. Boycott KM, Hartley T, Kernohan KD, et al. Care4Rare Canada: Outcomes from a decade of network science for rare disease gene discovery. *Am J Hum Genet*. 2022;109(11):1947-1959.

156. Wright CF, Fitzgerald TW, Jones WD, et al. Genetic diagnosis of developmental disorders in the DDD study: a scalable analysis of genome-wide research data. *Lancet*. 2015;385(9975):1305-1314.

157. Kim SY, Lee S, Woo H, et al. The Korean undiagnosed diseases program phase I: expansion of the nationwide network and the development of long-term infrastructure. *Orphanet J Rare Dis*. 2022;17(1):372.

158. Baynam G, Pachter N, McKenzie F, et al. The rare and undiagnosed diseases diagnostic service - application of massively parallel sequencing in a state-wide clinical service. *Orphanet J Rare Dis*. 2016;11(1):77.

159. Takahashi Y, Mizusawa H. Initiative on Rare and Undiagnosed Disease in Japan. *JMA J*. 2021;4(2):112-118.

160. Lochmüller H, Badowska DM, Thompson R, et al. RD-Connect, NeurOmics and EURenOmics: collaborative European initiative for rare diseases. *Eur J Hum Genet*. 2018;26(6):778-785.

161. López-Martín E, Martínez-Delgado B, Bermejo-Sánchez E, Alonso J, SpainUDP Network, Posada M. SpainUDP: The Spanish Undiagnosed Rare Diseases Program. *Int J Environ Res Public Health*. 2018;15(8). doi:10.3390/ijerph15081746

162. Salvatore M, Polizzi A, De Stefano MC, et al. Improving diagnosis for rare diseases: the experience of the Italian undiagnosed Rare diseases network. *Ital J Pediatr*. 2020;46(1):130.